\documentclass[final,1p,times]{elsarticle} 
\usepackage{graphicx} 
\usepackage{amssymb} 
\usepackage{amsthm} 
\usepackage{lineno} 
\newcommand{\sqrtsNN}{\mbox{$\sqrt{\mathrm{s}_{_{\mathrm{NN}}}}$} }
\newcommand{\vtwo}{$v_{2}$ }

\newcommand{\ptt}{$p_{T}$ }

\newcommand{\ks}{$\mathrm{K}^{0}_{S}$ }

\newcommand{\lam}{$\Lambda$ }

\newcommand{\phimeson}{$\phi$ }

\newcommand{\xii}{$\Xi$ }

\newcommand{\om}{$\Omega$ }

\def \GeVc {\mbox{$\mathrm{GeV}/c$}}

\def \auau  {$\mathrm{Au + Au}$ }
\def \cucu  {$\mathrm{Cu + Cu}$ }
\def \pp    {$p + p$ }

\journal{Nuclear Physics A} 
\begin{document} 

\begin{frontmatter} 


\title{Event anisotropy $v_2$ at STAR}

\author{Shusu Shi$^{a, b, c}$ for the STAR Collaboration}

\address[a]{Nuclear Science Division, Lawrence Berkeley National
Laboratory, Berkeley, CA, 94720, USA}
\address[b]{Institute of Particle Physics, Huazhong
Normal University, Wuhan, Hubei, 430079, China}
\address[b]{The Key Laboratory of Quark and Lepton Physics (Huazhong Normal
University), Ministry of Education, Wuhan, Hubei, 430079, China}

\begin{abstract} 
Collective flow reflects the dynamical evolution in high-energy
heavy ion collisions. In particular, the elliptic flow reflects
early collision dynamics~\cite{STARWhite}. We present a systematic
analysis of elliptic flow (\vtwo) for identified particles measured
in \auau and \cucu collisions at \sqrtsNN = 200 GeV. Number of quark
scaling is tested in the intermediate $p_{T}$ region and in the
smaller system (\cucu). The \cucu collisions results are compared
with those from ideal hydrodynamic model calculations.
\end{abstract} 

\end{frontmatter} 



\section{Introduction}
When two nuclei collide in non-central heavy-ion collisions, their
overlap area in the transverse plane has a short axis, which is
parallel to the impact parameter, and a long axis, which is
perpendicular to it. This initial spatial anisotropy of the overlap
region of the colliding nuclei is transformed into an anisotropy in
momentum space through interactions between the particles. The
magnitude of this effect is characterized by elliptic flow, defined
as
\begin{equation} v_{2}=\langle\cos2(\phi-\Psi_{R})\rangle
\end{equation}where $\phi$ is azimuthal angle of an outgoing particle,
$\Psi_{R}$ is the azimuthal angle of the impact parameter, and
angular brackets denote an average over many particles and events.

The characterization of the elliptic flow of produced particles by
their azimuthal anisotropy has proven to be one of the more fruitful
probes of the dynamics in \auau collisions at the Relativistic Heavy
Ion Collider(RHIC)~\cite{flow1,flow2,flow3}, see recent review
in~\cite{review1, review2, review3}. It can provide much information
about the degree of thermalization of the hot and dense medium. A
systematic study of the $p_{T}$ dependence of $v_{2}$ for different
particle species enables investigation of underlying phenomena and
the properties of the produced matter.

\section{Methods and Analysis}
In this proceeding, we report \vtwo measurements by the STAR
experiment from \sqrtsNN = 200 GeV \auau and \cucu collisions. Data
were taken from Run 5 (2005) and Run 7 (2007). STAR's Time
Projection Chamber (TPC)~\cite{STARtpc} is used as the main detector
for particle identification and event plane determination. The
centrality was determined by the number of tracks from the pseudorapidity region
$|\eta|\le 0.5$. Two Forward Time Projection Chambers (FTPCs) were also
used for event plane determinations. The FTPCs cover 2.5 $\le |\eta|
\le$ 4. The pseudorapidity gap between FTPC and TPC allows us to
reduce some of the non-flow effects. In \auau collisions, the
difference between $v_{2}$(TPC) (event plane determined by TPC
tracks) and $v_{2}$(FTPC) (event plane determined by FTPC tracks)
has been used to estimate the systematic errors, where in \cucu
collisions, we used $v_{2}$(FTPC) for the measurement,
$v_{2}^{AA-pp}$(FTPC)~\cite{AA-pp} (subtracting the residual non-flow effects based on the azimuthal
correlations in \pp collisions) 
for the systematic study.

 The PID is achieved via dE/dx in TPC and topologically reconstructed hadrons:
\ks $\rightarrow \pi^{+} + \pi^{-}$, $\phi \rightarrow K^{+} +
K^{-}$, \lam $\rightarrow p + \pi^{-}$ ($\overline{\Lambda}
\rightarrow \overline{p} + \pi^{+}$),
 $\Xi^{-} \rightarrow$ \lam $+\ \pi^{-}$ ($\overline{\Xi}^{+}
\rightarrow$ $\overline{\Lambda}$+\ $\pi^{+}$) and $\Omega^{-}
\rightarrow$ \lam $+\ K^{-}$ ($\overline{\Omega}^{+} \rightarrow$
$\overline{\Lambda}$+\ $K^{+}$). The detailed description of the
procedure can be found in Refs.~\cite{flow4, klv2_130GeV,starklv2}.

\section{Results and Discussions}

\begin{figure*}[ht]
\centering \hskip -.0cm \vskip -.0cm
\includegraphics[totalheight=0.4\textheight]{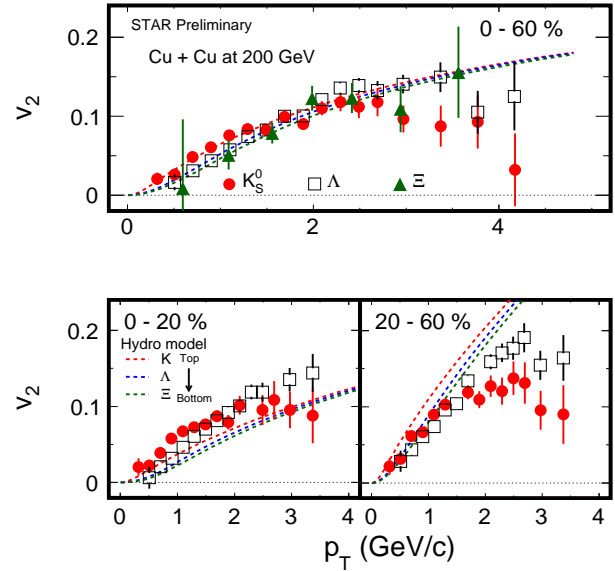}
\caption{The \vtwo as a function of \ptt for \ks , \lam and \xii in
0 - 60 \% (top), 0 - 20 \% and 20 - 60 \% (bottom) \cucu collisions
at \sqrtsNN = 200 GeV. Dashed lines represent ideal hydrodynamical
calculation~\cite{hydro_pasi}.} \label{fig_pid}
\end{figure*}

Results using ideal hydrodynamical calculations~\cite{hydro_pasi1}
have been able to reproduce mass ordering of \vtwo in the low \ptt
region in Au+Au collisions. Figure 1 shows the \vtwo for \ks , \lam
and \xii as a function of \ptt in different centrality selections
for \cucu collisions along with results of hydrodynamical
calculations~\cite{hydro_pasi}. We observe that \vtwo for \lam is
smaller than \vtwo for \ks for \ptt $<$ 2 GeV/c. For \ptt $>$ 2
GeV/c, \vtwo for \lam becomes larger than that of \ks. We have also
found \xii has sizable \vtwo in minimum bias 0 - 60 \% centrality.
The ideal hydrodynamical model does not describe the centrality
dependence of our data. For 0 - 20 \%, the model under-predicts the
data and for 20 - 60 \%, it over-predicts the \vtwo. Effects not
included in the model which may be relevant are geometrical
fluctuations in the initial conditions (particularly important in
central collisions), finite viscosity effects and incomplete
thermalization. It remains to be seen if these effects can account
for the difference between the models and data.

\begin{figure*}[ht]
\centering \hskip -.0cm \vskip .0cm
\includegraphics[totalheight=0.3\textheight]{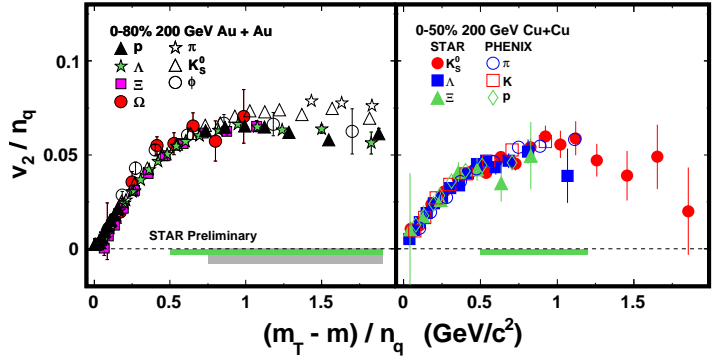}
\caption{Number of quark scaling of \vtwo as a function of $m_{T} -
mass$ in 0 - 80 \% \auau (left) and 0 - 50 \% \cucu (right)
collisions at \sqrtsNN = 200 GeV. Green and gray bands show non-flow
systematic errors for strange hadrons ($(m_{T} - mass)/n_{q} > 0.5
 GeV/c^{2}$) and $\pi$, p ($(m_{T} - mass)/n_{q} > 0.75
 GeV/c^{2}$) respectively. PHENIX results were taken
from~\cite{PHENIXcucu}.} \label{fig_pid}
\end{figure*}

Quark coalescence~\cite{cola} or recombination~\cite{recom} mechanisms in
particle production predict that at intermediate $p_{T}$ (2 \GeVc $<p_{T}<$ 5 \GeVc) number of
quark (NQ) scaled \vtwo will follow a universal curve. Thus, the NQ
scaling is considered evidence for partonic degrees of freedom in Au
+ Au collisions at \sqrtsNN = 200 GeV~\cite{starklv2}. With the
large statistics from Run 7, we can test the scaling in the large
$p_{T}$ region. Figure 2 shows the number of quark scaled \vtwo for
identified particles as a function of $(m_{T} - mass)/n_{q}$ in
\auau and \cucu collisions at \sqrtsNN = 200 GeV. Proton and \lam begin
to deviate from the NQ scaling when $(m_{T} - mass)/n_{q} >
1~GeV/c^{2}$ in the \auau case. Scaling behavior can be seen in the smaller
system (\cucu) at the same energy.

\begin{figure*}[ht]
\centering \hskip -0.5cm \vskip 0cm
\includegraphics[totalheight=0.3\textheight]{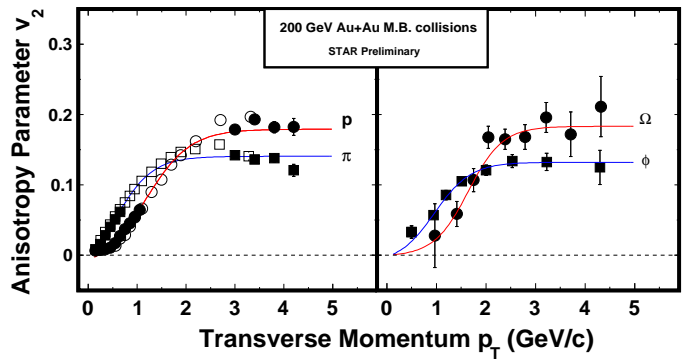}
\caption{\vtwo as function of $p_{T}$ for $\pi$, $p$ (left) and
$\phi$, \om (right) in Au + Au minimum-bias  collisions at \sqrtsNN
= 200 GeV. Open symbols represent results from
PHENIX~\cite{PHENIXpip}. Lines represent NQ-inspired
fit~\cite{xinv2}.} \label{fig_pid}
\end{figure*}

Figure 3 shows the \vtwo for \phimeson and \om together with \vtwo
for $\pi$ and $p$ as a function of \ptt in minimum bias \auau
collisions at \sqrtsNN = 200 GeV. The $p_{T}$ dependence of \vtwo
for $\pi$ and $p$ is observed to be similar as the corresponding
results for $\Omega$ baryons and $\phi$ mesons. This indicates that the heavier s
quarks flow as strongly as the lighter u and d quarks, providing
evidence for partonic collectivity.

\section{Summary}
In summary, we present the results from a systematic analysis of the
identified particles elliptic flow ($v_2$) measurement  from \auau
and \cucu collisions at \sqrtsNN = 200 GeV. Ideal hydrodynamic model
calculations fail to reproduce the data in \cucu collisions. Proton and \lam begin
to deviate from the NQ scaling when $(m_{T} - mass)/n_{q} >
1~GeV/c^{2}$ in \auau collisions; scaling behavior can be seen in the smaller
system (\cucu). The fact that the \phimeson and \om \vtwo
($p_{T}$) follows a similar trend as that of  $\pi$ and $p$
indicates that the heavier s quarks flow as strongly as the lighter
u and d quarks suggesting partonic collectivity has been established
at RHIC .

\section{Acknowledgments}
The author was supported in part by the National Natural Science
Foundation of China under grant no. 10775058, MOE of China under
project IRT0624 and MOST of China under Grant No 2008CB817707.


\begin{thebibliography}{10}
\bibitem{STARWhite} STAR Collaboration, J. Adams et al., Nucl. Phys. {\bf A 757},
(2005)102.
\bibitem{flow1} J. Adams et al., (STAR Collaboration), Phys. Rev. {\bf C 72}, 014904 (2005).
\bibitem{flow2} S.S. Adler et al., (PHENIX Collaboration), Phys. Rev. Lett. {\bf 94}, 232302 (2005).
\bibitem{flow3} B.B. Back et al., (PHOBOS Collaboration), Phys. Rev. {\bf C 72}, 051901 (2005).
\bibitem{review1} S.A. Voloshin, A.M. Poskanzer and R. Snellings,
arXiv:nucl-ex/0809.2949.
\bibitem{review2} P.~Sorensen, arXiv:nucl-ex/0905.0174.
\bibitem{review3} D. Teaney, arXiv:nucl-th/0905.2433.
\bibitem{STARtpc} K.~H.~Ackermann {\it et al.} (STAR
Collaboration), Nucl. Instrum. Methods A {\bf 499}, 624 (2003).
\bibitem{AA-pp} J. Adams et al., (STAR Collaboration), Phys. Rev. Lett. {\bf 93}, 252031 (2004).
\bibitem{flow4} B.I. Abelev et al., (STAR Collaboration), Phys. Rev. {\bf C 77}, 054901 (2008).
\bibitem{klv2_130GeV} C.~Adler {\it et al.} (STAR Collaboration),
Phys. Rev. Lett. {\bf 89}, 132301 (2002).
\bibitem{starklv2} J. Adams {\it et al.} (STAR Collaboration),
Phys. Rev. Lett. {\bf 92}, 052302 (2004).
\bibitem{hydro_pasi1} P. Huovinen, Private communication, 2006.
\bibitem{hydro_pasi} P. Huovinen, Private communication, 2008.

\bibitem{cola}D. Molnar and S. A.Voloshin, Phys. Rev. Lett. {\bf 91}, 092301 (2003).

\bibitem{recom} R.C. Hwa and C. B. Yang, Phys. Rev. {\bf C 67}, 064902 (2003); R. J.
Fries, B. Muller, C. Nonaka, and S. A. Bass, Phys. Rev. Lett. {\bf
90}, 202303 (2003).


\bibitem{PHENIXpip}PHENIX Collaboration, M. Issah and A. Taranenko,
(2006), nucl-ex/0604011.
\bibitem{xinv2} X. Dong et al., Phys. Lett. B {\bf 597}, 328 (2004).
\bibitem{PHENIXcucu}PHENIX Collaboration, M. Issah, arXiv:0805.4039.









\end{thebibliography}
\end{document}